\documentclass[superscriptaddress,twocolumn,showpacs,amsmath,amssymb]{revtex4}
\usepackage{graphicx}

\renewcommand{\vec}[1]{\mbox{\boldmath $#1$}}

\begin{document}

\title{
Correlated two-neutron emission in the decay of unbound nucleus $^{26}$O}

\author{K. Hagino}
\affiliation{ 
Department of Physics, Tohoku University, Sendai 980-8578,  Japan} 

\author{H. Sagawa}
\affiliation{
Center for Mathematics and Physics,  University of Aizu, 
Aizu-Wakamatsu, Fukushima 965-8560,  Japan}
\affiliation{
RIKEN Nishina Center, Wako 351-0198, Japan}


\begin{abstract}
The particle unbound $^{26}$O nucleus 
is located outside the neutron drip line, and 
spontaneously decays by emitting two neutrons with a relatively 
long life time due to the centrifugal barrier.
We study the decay of this nucleus with 
a three-body model 
assuming an inert $^{24}$O core and two valence neutrons. 
We first point out 
the importance of the neutron-neutron final state interaction 
in the observed decay energy spectrum. 
We also show that the energy and 
 and angular distributions for the 
two emitted neutrons manifest a clear evidence for the strong 
neutron-neutron correlation in the three-body resonance state. 
In particular, we find an enhancement 
of two-neutron emission in back-to-back directions. 
This is interpreted as 
a consequence of {\it dineutron correlation}, with which the two 
neutrons are spatially localized before the emission. 
\end{abstract}

\pacs{21.10.Tg,23.90.+w,25.60.-t,21.45.-v}

\maketitle

Correlations among particles lead to a variety of rich phenomena in 
many-fermion systems, such as superconductivity and superfluidity. 
The spatial distribution of particles is also affected by 
the correlations. For many-electron systems, the Coulomb repulsion 
between electrons yields the so called Coulomb hole, in which the 
distribution of the second electron is largely suppressed in the 
vicinity of the first electron \cite{CN61,RRB78}. 
In atomic nuclei, in contrast, an attractive nuclear force leads 
to the dineutron and diproton correlations, with 
which two nucleons are spatially localized in the surface region of 
nuclei\cite{BBR67,CIMV84}. These nuclear correlations 
have attracted lots of attention recently 
\cite{BE91,Zhukov93,HS05,MMS05,PSS07}, 
in connection to physics of weakly bound nuclei. 

In order to probe the inter-particle 
correlation, it has been a standard way 
in atomic physics to measure a double ionization with strong laser 
fields\cite{WSD94,WGW00,BKJ12,BLHE12}. It has been observed that 
the ionization rate 
is significantly enhanced due to the electronic correlation, and moreover, 
there is a strong momentum correlation between the two emitted electrons. 
The corresponding experiment in nuclear physics is the Coulomb breakup 
of the Borromean nuclei $^{11}$Li and $^6$He, in which those nuclei 
are broken up to the core nuclei, $^9$Li and $^4$He, and two neutrons 
in the Coulomb field of a target nucleus \cite{N06,A99,NK12}. 
The observed breakup probabilities, 
especially those for the $^{11}$Li nucleus, 
show a sharp peak in the low-energy 
region, which can be accounted for only by taking 
into account the neutron-neutron correlations. 
Furthermore, from the observed strength distribution, the opening angle 
between the valence neutrons in the ground state of the Borromean nuclei 
has been inferred employing the cluster sum rule \cite{N06,HS07,BH07}. 
For both $^{11}$Li and $^6$He, 
the extracted opening angles were significantly smaller  
than the value 
for the independent neutrons, that is, 
90 degrees,  and clearly indicate the existence of the dineutron 
correlation. 

A small drawback with the cluster sum rule approach is that it yields only an 
expectation value of the opening angle and 
a detailed angular 
distribution cannot be studied with this method. For this reason, 
the energy and the angular distributions of the emitted neutrons from 
the Coulomb breakup have been investigated\cite{EB92}. However, 
it has been concluded that those distributions are largely determined 
by the properties of the neutron-core system, and thus it is difficult 
to acquire detailed information on the neutron-neutron 
correlations 
from 
the Coulomb breakup measurement \cite{HSNS09,KKM10}. 

It is therefore desirable to seek for other probes for the 
nucleonic correlation. Among them, the two-proton radioactivity, that is, 
the spontaneous emission of two protons of proton-unbound nuclei, has been 
considered to be a good candidate for that purpose \cite{PKGR12}. 
An attractive feature of this phenomenon is that the two valence protons 
are emitted without an influence of disturbance of nuclei due to 
an external field. 
Very recently, the ground state {\it two-neutron} emission
was discovered for $^{16}$Be\cite{SKB12}.
Earlier measurements on the two-neutron emission 
include those for $^{10}$He \cite{J10} and $^{13}$Li \cite{J10,A08}. 
These are a counter part of the two-proton emission of proton-rich nuclei, 
corresponding to a penetration of two neutrons over a centrifugal
barrier. 
Subsequently, the two-neutron emission was discovered also for 
$^{26}$O\cite{LDK12,CSA12} and $^{13}$Li \cite{KLD13}. 
So far, the experimental data have been analyzed 
only with a schematic dineutron model \cite{SKB12,KLD13} 
(see also Ref. \cite{MOA12}).  
Although such schematic model appears to reproduce 
the data, 
realistic three-body model calculations with configuration mixings 
and full neutron-neutron correlations 
have been clearly urged. 

In this paper, we apply the three-body model with a density-dependent 
contact interaction between the valence neutrons to the decay problem of 
$^{26}$O, assuming $^{24}$O to be an inert core.  
This model 
has been successfully applied to describe the ground state properties 
and the {Coulomb break-up} of neutron-rich nuclei\cite{BE91,HS05,HSNS09,EBH97}. 
In order to describe the decay of neutron-unbound 
nucleus, we shall take into account the couplings to 
continuum by the Green's function technique, 
which was invented 
in Ref. \cite{EB92} in order to describe 
the continuum dipole excitations 
of $^{11}$Li. 
We shall 
discuss the role of neutron-neutron correlation in the decay probability, 
as well as in 
the energy and the angular distributions of the emitted neutrons. 

In the experiment of Ref. \cite{LDK12}, the $^{26}$O nucleus was produced 
in the 
single proton-knockout reaction from a secondary $^{27}$F beam. 
We therefore first construct the ground state of $^{27}$F with a three-body 
model, assuming the $^{25}$F+$n$+$n$ structure. 
We then assume a sudden proton removal, that is, the $^{25}$F  core  
changes 
 to $^{24}$O  keeping the configuration 
for the $n$+$n$ subsystem of $^{26}$O to be the same as in the ground state of $^{27}$F. 
This initial state, $\Psi_i$, 
is then evolved with the Hamiltonian for the three-body 
$^{24}$O+$n$+$n$ system 
for the two-neutron decay. 

We therefore consider two three-body Hamiltonians, one for the initial state
$^{25}$F+$n$+$n$ and the other for the final state $^{24}$O+$n$+$n$.  
For both the systems, we use similar Hamiltonians as 
that in Refs. \cite{HS05,EBH97}, 
\begin{equation}
H=\hat{h}_{nC}(1)+\hat{h}_{nC}(2)+v(1,2)
+\frac{\vec{p}_1\cdot\vec{p}_2}{A_cm}, 
\label{3bh}
\end{equation}
where $A_c$ is the mass number of the core nucleus, 
$m$ is the nucleon mass, 
and $\hat{h}_{nC}$ is the single-particle (s.p.) Hamiltonian for a valence
neutron interacting with the core. 
The last term in
Eq. (\ref{3bh}) is the two-body part of the 
recoil kinetic energy of the core nucleus \cite{EBH97}, while the 
one-body part is included in 
$\hat{h}_{nC}$. 
We use a contact interaction 
between the valence 
neutrons, $v$, given as\cite{BE91,HS05,EBH97}, 
\begin{equation}
v(\vec{r}_1,\vec{r}_2)=\delta(\vec{r}_1-\vec{r}_2)
\left(v_0+\frac{v_\rho}{1+\exp[(r_1-R_\rho)/a_\rho]}\right). 
\label{vnn}
\end{equation}
Here, the strength $v_0$ is determined 
to be $-$857.2 MeV$\cdot$fm$^{3}$
from 
the scattering length for the 
$nn$ scattering together with the cutoff energy, which we take 
$E_{\rm cut}=30$ MeV. 
See Refs.\cite{HS05,EBH97} for the details. 
The second term in Eq. (\ref{vnn}) 
simulates the density dependence of the interaction. 
Taking 
$R_\rho=1.34\times A_c^{1/3}$ fm and $a_\rho$=0.72 fm, 
we adjust the value of $v_\rho$ to be 952.3 MeV$\cdot$fm$^{3}$
so as to reproduce the experimental two-neutron separation 
energy of $^{27}$F, $S_{\rm 2n}$=2.80(18) MeV\cite{JSM07}. 

We employ a Woods-Saxon form for the s.p. potential in $\hat{h}_{nC}$. 
For the $^{24}$O+$n+n$ system, 
we 
take $a=0.72$ fm and 
$R_0=1.25A_c^{1/3}$ fm with $A_c=24$, and determine  
the values of 
$V_0=-44.1$ MeV and $V_{ls}$=45.87 MeV$\cdot$fm$^2$ in order to reproduce 
the single-particle energies of 
$\epsilon_{2s_{1/2}}=-4.09(13)$ MeV and 
$\epsilon_{1d_{3/2}}=770^{+20}_{-10}$ keV \cite{H08}. 
This potential yields 
the width for the 1$d_{3/2}$ state of 
$\Gamma_{1d_{3/2}}=92.9$ keV, which is compared with 
the empirical value, 
$\Gamma_{1d_{3/2}}=172(30)$ keV \cite{H08}. 
For the $^{25}$F+$n+n$ system, 
one has to modify the Woods-Saxon potential in order to take into account 
the presence of the valence proton in the core nucleus. 
The important effect comes from the tensor force 
between the valence proton and neutrons \cite{O05}, 
which primarily modifies the spin-orbit potential 
in the mean-field approximation\cite{SBF77,CSFB07,LBB07}. 
We thus use the same Woods-Saxon potential for 
$^{25}$F+$n+n$ system 
as that for the $^{24}$O+$n+n$ system except for the spin-orbit 
potential, whose strength is weakened 
to $V_{ls}$=33.50 MeV$\cdot$fm$^2$ 
in order to reproduce 
the energy of 
$1d_{3/2}$ state in $^{27}$F, 
$\epsilon_{1d_{3/2}}=-0.811$ MeV. 

With the initial wave function thus obtained, the decay energy spectrum 
can be computed as \cite{EB92}, 
\begin{eqnarray}
\frac{dP}{dE}&=&\frac{1}{\pi}\Im \langle \Psi_i|G_0(E)|\Psi_i\rangle \nonumber \\
&&-\frac{1}{\pi}
\Im\langle \Psi_i|G_0(E)v(1+G_0(E)v)^{-1}G_0(E)|\Psi_i\rangle.
\label{decayenergy}
\end{eqnarray}
where $\Im$ denotes the imaginary part.  In Eq. \eqref{decayenergy}, $G_0(E)$ is the unperturbed 
Green's function given by,
\begin{equation}
G_0(E)=\sum_{\rm 1,2}\frac{|(j_1j_2)^{(0^+)}\rangle\langle(j_1j_2)^{(0^+)}|}
{e_1+e_2-E-i\eta},
\label{green0}
\end{equation}
where $\eta$ is an infinitesimal number and 
the sum includes all independent two-particle states coupled to the 
total angular momentum of $J=0$ with the positive parity, described by the 
three-body Hamiltonian for $^{24}$O+$n+n$. 
As in our previous study for the continuum E1 excitations of the $^{11}$Li 
nucleus \cite{HSNS09}, 
we have neglected 
the two-body part of the recoil kinetic energy 
in order to derive Eq. (\ref{decayenergy}), while we keep all 
the recoil terms in constructing the initial state wave function. 

\begin{figure} 
\includegraphics[scale=0.35,clip]{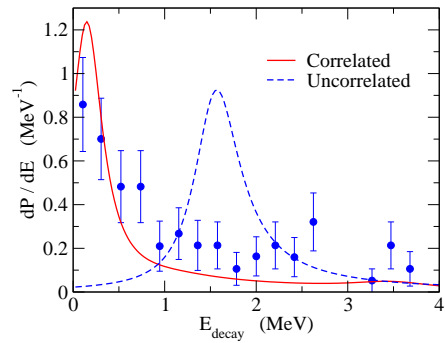}
\caption{(Color online) 
The decay energy spectrum for the two-neutron emission decay of $^{26}$O. 
The solid line denotes the result with the full inclusion of the final 
state neutron-neutron ($nn$) interaction, while the dashed line shows the 
result without the final state $nn$ interaction. 
The theoretical curves are drawn with a finite width of 0.21 MeV,  
which is the same as the experimental energy resolution. 
The experimental data, 
normalized to the unit area, are 
taken from Ref. \cite{LDK12}. 
}
\end{figure}

Figure 1 shows the decay energy spectrum obtained with 
Eq. \eqref{decayenergy}. 
The solid line shows the correlated spectrum, in which the final state $nn$ 
interaction is fully taken into account, while the dashed line shows the 
result without the final state $nn$ interaction. 
The latter corresponds to the first term in Eq. (\ref{decayenergy}). 
Since the width of the three-body resonance state is extremely small, 
which is 
experimentally 
the order of 10$^{-10}$ MeV \cite{K13}, we have 
introduced a finite width for a presentation purpose. That is, in evaluating 
the unperturbed Green's function, Eq. (\ref{green0}), we set  $\eta$ =0.21 MeV, 
that is to be the same as the experimental energy 
resolution. 
Without the final state $nn$ interaction, 
the two valence neutrons 
in $^{26}$O 
occupy the s.p. resonance state of 
1$d_{3/2}$ at 770 keV, and the peak in the decay energy spectrum appears 
at twice this energy. When the final state $nn$ interaction is taken  
into account, the peak is largely shifted towards a lower energy and 
appears at 0.14 MeV, in a good agreement with the experimental data. 

\begin{figure} 
\vspace{-0.9cm}
\includegraphics[scale=1.,clip]{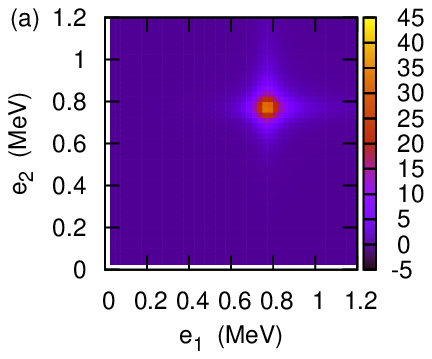}\\
\vspace{-0.9cm}
\includegraphics[scale=1.,clip]{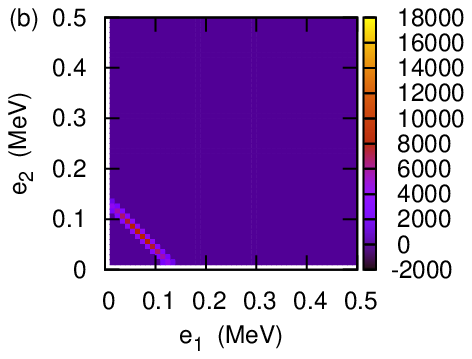}
\caption{(Color online) 
The decay probability distribution for the two-neutron emission decay of 
$^{26}$O as a function of the energies of the two emitted neutrons. 
Fig. 2(b) shows the correlated probability while Fig. 2(a) 
shows the uncorrelated probability without the final state $nn$ interactions. 
}
\end{figure}

The energy distribution of the two emitted neutrons is shown in Fig. 2, 
in which 
a decay amplitude is calculated to a specific 
two-particle final state \cite{EB92}, 
\begin{eqnarray}
M_{j,l,k_1,k_2}
&=&\langle (jj)^{(00)}|1-vG_0+vG_0vG_0-\cdots|\Psi_i\rangle, 
\label{amplitude1}
\\
&=&\langle (jj)^{(00)}|(1+vG_0)^{-1}|\Psi_i\rangle.
\label{amplitude2}
\end{eqnarray}
The unperturbed Green's function, $G_0$, 
is evaluated at $E=e_1+e_2$. Notice that a series of 
$-vG_0+vG_0vG_0-\cdots$ in Eq. (\ref{amplitude1}) describes
the multiple  rescattering effect of 
the two neutrons during the emission due to the final state $nn$ 
interaction, which is included to the all orders in Eq. (\ref{amplitude2}). 
In contrast to the case of decay energy spectra shown 
in Fig. 1,   we take $\eta$ 
in Eq. (\ref{green0}) to be an infinitesimal number 
in evaluating 
the unperturbed Green's function 
and use the Gauss-Legendre 
integration technique for  Eq. \eqref{amplitude2} 
as described in Ref. \cite{EB92}. 
The energy spectrum is then computed as,
\begin{equation}
\frac{d^2P}{de_1de_2}=\sum_{j,l}|M_{j,l,k_1,k_2}|^2\,\frac{dk_1}{de_1}
\frac{dk_2}{de_2}, 
\end{equation}
where the factors $dk/de$ are due to the normalization of the continuum 
single-particle wave functions, for which we follow Ref. \cite{EB92}. 

Figure 2(a) shows the energy distribution obtained by switching off 
the final 
state $nn$ interaction. 
The energy distribution is dominated by the single-particle $d_{3/2}$ 
resonance state at 0.77 MeV. 
A ridge appears as in the energy 
distribution for dipole excitations of Borromean nuclei \cite{EB92,HSNS09}. 
The energy distribution with the $nn$ final state interaction is shown 
in Fig. 2(b). The energy distribution is drastically changed, being 
highly concentrated along 
the line of $e_1+e_2\sim$ 0.14 MeV with an extremely small width. 
The variation with $e_1$ is weak 
along this line, although the maximum still appears at $e_1=e_2$. 
This is a clear manifestation of a three-body resonance, and is in 
marked contrast to the continuum dipole excitations, in which 
the final state $nn$ interaction does not affect much the 
shape of the energy distribution \cite{HSNS09}. 
\begin{figure} 
\includegraphics[scale=0.35,clip]{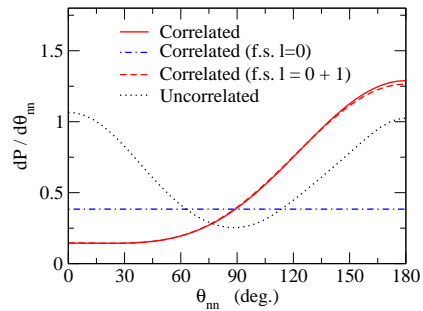}
\caption{(Color online) 
The differential probability distribution with respect to the opening angle of the emitted 
two neutrons from 
 $^{26}$O. 
The solid and the dotted lines show the correlated and uncorrelated results, 
respectively. The dot-dashed and the dashed lines denote the correlated 
results obtained by including the angular momentum of the final state 
up to $l=0$ and $l=1$, respectively. }
\end{figure}

The angular distribution of the emitted neutrons can be also calculated 
 using 
the decay amplitude, Eq. (\ref{amplitude1}). 
The amplitude for emitting the 
two neutrons with spin components of $s_1$ and $s_2$ and momenta 
$\vec{k}_1$ and $\vec{k}_2$ reads \cite{EB92}, 
\begin{eqnarray}
f_{s_1s_2}(\vec{k}_1,\vec{k}_2)&=&
\sum_{j,l}e^{-il\pi}e^{i(\delta_1+\delta_2)}\,
M_{j,l,k_1,k_2} \nonumber \\
&&\times \langle [{\cal Y}_{jl}(\hat{\vec{k}}_1)
{\cal Y}_{jl}(\hat{\vec{k}}_2)]^{(00)}|\chi_{s_1}\chi_{s_2}\rangle,
\label{angularamplitude}
\end{eqnarray}
where ${\cal Y}_{jlm}$ is the spin-spherical harmonics, 
$\chi_s$ is the spin wave function, and $\delta$ is the nuclear 
phase shift.
The angular distribution is then obtained as 
\begin{equation}
\frac{dP}{d\theta_{12}}=4\pi\sum_{s_1,s_2}
\int dk_1dk_2\, |f_{s_1s_2}(k_1,\hat{\vec{k}}_1=0,k_2,
\hat{\vec{k}}_{2}=\theta_{12})|^2,
\label{angular}
\end{equation}
where we have set $z$-axis to be parallel to $\vec{k}_1$ and 
evaluated the angular distribution as a function of the 
opening angle, $\theta_{12}$,  
of the two emitted neutrons. 

The angular distribution obtained without including the final state 
$nn$ interaction is shown by the dotted line in Fig. 3. 
The main component in the initial wave function, $\Psi_i$, is the 
$d_{3/2}$ configuration, and the angular distribution is almost symmetric 
around $\theta_{12}=\pi/2$. 
In the presence of the final state $nn$ interaction, the angular distribution 
becomes highly asymmetric, in which the emission of two neutrons in the 
opposite direction (that is, $\theta_{12}=\pi$) is 
enhanced\cite{GMZ13}, as is shown 
by the solid line. 
Notice that we have obtained the correlated distribution 
by evaluating Eq. (\ref{angular}) 
only at $e_1+e_2=0.14$ and then normalize it, since it is hard 
to carry out the 
integrations in Eq. (\ref{angular}) when the resonance width 
is extremely small. We do not expect that this procedure causes any 
significant error in evaluating the angular distribution. 
The asymmetric angular distribution for the correlated case 
originates from the interference 
between opposite party components, as in the dineutron correlation 
in the density distribution \cite{CIMV84}. 
For the $^{26}$O nucleus, it is due to the interference between the 
$l=0$ and $l=1$ 
components. The dot-dashed line in Fig. 3 shows 
the result obtained by including only $l=0$ in Eq. (\ref{angularamplitude}), 
while 
the dashed line shows the result with $l$=0 and 1. 
One can see that the angular distribution is almost exhausted by  
these two angular momenta and they contribute with almost equal amplitudes.  
For higher partial waves $l\geq2$, 
the scattering wave functions in Eq. (\ref{amplitude2}) are highly 
damped inside the centrifugal barrier since the energy is quite low 
($e_1\sim e_2 \sim$ 0.07 MeV).  In other words, 
the two neutrons 
are rescattered into $s$-wave and $p$-wave states by multistep process due to the interaction $v$ 
(see Eq. (\ref{amplitude1})) 
and these low $l$ components uniquely enhance the penetrability, even though the main 
component in the initial wave function is the $d$-wave state. 
This picture is consistent with what Grigorenko {\it et al.} have argued 
in Ref.\cite{GMZ13}. 

\begin{figure} 
\includegraphics[scale=1.,clip]{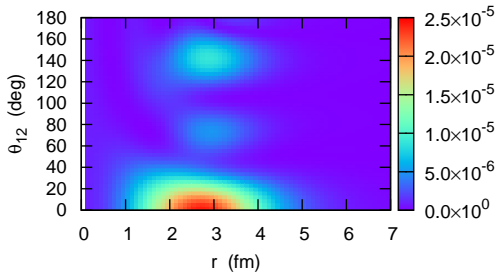}
\caption{(Color online) 
The two-particle density for the resonance state of $^{26}$O 
obtained with the box boundary condition. It is plotted as a 
function of $r_1=r_2=r$ and the angle between the valence neutrons, 
$\theta_{12}$. 
}
\end{figure}

The enhancement of angular distribution at backward angles for $^{26}$O 
has also been seen theoretically in the dipole 
excitations of $^{11}$Li \cite{EB92} and 
both theoretically and experimentally  
in the two-proton emission decay 
of $^6$Be \cite{G09}. This reflects the spatial correlation of the 
three-body resonance state of $^{26}$O. 
Figure 4 shows 
the two-particle density for a resonance state of $^{26}$O obtained with 
the box boundary condition as a 
function of $r_1=r_2=r$ and the opening angle between 
the two neutron, $\theta_{12}$. 
One finds that the density distribution is 
well localized in the small $\theta_{12}$ region, which is clear manifestation 
of the dineutron correlation \cite{HS05}. 
It has been well known that the configurations with opposite parity 
have to contribute coherently in order to form the dineutron 
correlation \cite{CIMV84,PSS07,HVPBS11}. 
In the angular distribution in Fig. 3, a phase factor, 
$e^{-il\pi}$, in the amplitude in Eq. (\ref{angularamplitude}) 
alters the sign of the contributions of odd partial waves, leading 
to the opposite tendency from the density, that is, 
the preference of emission of two-neutrons 
in the back-to-back angles. 
The nuclear phase shifts, $\delta_1+\delta_2$, plays a minor role in 
the decay of $^{26}$O, partly 
because the decay energy is extremely small. 
Evidently, the back-to-back emission of two neutrons 
in the momentum space
from the decay of 
$^{26}$O is another manifestation of 
the strong dineutron correlation in the coordinate space of ground state 
density distribution. 

For $^{16}$Be and $^{13}$Li, the experimental angular distributions show 
an enhancement of emission with relatively small 
opening angles\cite{SKB12,KLD13}. 
It has  yet to be clarified why 
these nuclei show different angular 
distributions from $^{26}$O (and from $^{6}$Be and $^{11}$Li). 
One possible reason is that the nuclear phase shift might play 
a more 
important role 
in these nuclei 
so that the phase factor 
$e^{-il\pi}$ is canceled out. 
Another reason may be the core excitation, with which 
the $nn$ configuration with coupled angular momenta of $J\neq0$ 
is largely admixed in the ground state wave function. 
In order to confirm these points, three-body model calculations for these 
nuclei with the core excitations are clearly needed, but 
we leave them as a future work. 

In summary, we have used the three-body model with a contact neutron-neutron 
interaction in order to analyze the two-neutron emission decay of the 
unbound neutron-rich nucleus $^{26}$O. 
Using the Green's function technique, we have analyzed the decay energy 
spectrum, the energy and the angular distributions of the two emitted 
neutrons.  We have pointed out 
that  the final state n-n interaction plays a crucial role to reproduce 
the strong low energy peak of the experimental decay energy spectrum. 
We have also argued that the energy distribution is  a clear 
manifestation of a three-body resonance state and its  density distribution is strongly reflected in 
the angular distribution of the emitted neutrons. 
In particular, the angular distribution clearly 
prefers the emission of the two neutrons in the back-to-back angles, 
that can be interpreted  as a clear evidence for the dineutron correlation. 
So far, the energy and the angular distributions for the two-neutron decay 
of $^{26}$O have not yet been measured experimentally. 
It would be extremely intriguing if they will be measured at 
new generation RI beam facilities, such as 
the SAMURAI facility at RIBF at RIKEN \cite{AN13}. 

\medskip

We thank Z. Kohley, T. Nakamura, A. Navin, Y. Kondo and T. Aumann for useful discussions. 
This work was supported by 
JSPS KAKENHI Grant Numbers 
22540262 and 25105503.

\end{document}